\documentclass[aip,jcp,reprint,twocolumn,showkeys]{revtex4-1}
\usepackage{amsmath}
\usepackage{bm}
\usepackage{bbm}
\usepackage{graphicx}
\usepackage{comment}
\usepackage{color}
\usepackage[utf8]{inputenc}

\usepackage[T1]{fontenc}
\definecolor{dred}{rgb}{.8,0.2,.2}
\definecolor{dblue}{rgb}{.2,0.2,.8}
\definecolor{green}{rgb}{0,0.65,0}
\newcommand{\ad}{a^\dagger}
\newcommand{\ket}[1]{|#1\rangle}
\newcommand{\bra}[1]{\langle#1|}
\newcommand{\Ff}{\hat{a}}
\newcommand{\unity}{\mathbbmss{1}}

\newcommand{\que}[1]{#1} 

\renewcommand{\eqref}[1]{Eq.~(\ref{#1})}

\usepackage{ulem}
\begin{document}
\normalem

\title{On the NP-completeness of the  Hartree-Fock method for translationally invariant systems}
\vspace*{10mm}
\author{James Daniel Whitfield}
\email{james.whitfield@univie.ac.at}
\affiliation{Vienna Center for Quantum Science and Technology, Department of Physics, University of Vienna, Boltzmanngasse 5, Vienna, Austria}

\author{Zolt\'an Zimbor\'as}
\email{z.zimboras@ucl.ac.uk}
\affiliation{Department of Computer Science, University College London, Gower Street, WC1E 6BT London, United Kingdom}
\affiliation{Department of Theoretical Physics, University of the Basque Country UPV/EHU, P.O. Box 644,  E-48080 Bilbao, Spain}

\date{\today}

\begin{abstract}
	The self-consistent field method utilized for solving the Hartree-Fock (HF) problem and the closely related Kohn-Sham problem, is typically thought of as one of the cheapest methods available to quantum chemists.  This intuition has been developed from the numerous applications of the self-consistent field method to a large variety of molecular systems.  However, as characterized by its  worst-case behavior, the HF problem is NP-complete. In this work, we map out boundaries of the NP-completeness by investigating restricted instances of HF.  We have constructed two new NP-complete variants of the problem.  The first is a set of Hamiltonians whose 
translationally invariant Hartree-Fock solutions are trivial, but whose broken symmetry solutions are NP-complete.  Second, we demonstrate how to embed instances of spin glasses into translationally invariant Hartree-Fock instances and provide a numerical example.  These findings are the first steps towards understanding in which cases the self-consistent field method is computationally feasible and when it is not.
\end{abstract}


\maketitle
\section{Introduction}
The Hartree-Fock (HF) method\cite{Hartree28,Fock30}  is one of the most important quantum chemistry techniques as it 
provides the mathematical setting for the chemist's notion of molecular 
orbitals widely used in organic chemistry. 
While this method is known to be a weak approximation in many cases (e.g.~in strongly correlated systems such as transition metal complexes, at non-equilibrium geometries, etc.),
it serves also the basis for more sophisticated (post-Hartree-Fock) algorithms which improve the mean-field wave function obtained.  Moreover, the self-consistent field (SCF) methodology used to finding the Hartree-Fock ground-state solution is also applied to solve Kohn-Sham density functional theory.  

For practitioners of quantum chemistry, the runtime of the Hartree-Fock
algorithm method is often dominated by the time needed to form and diagonalize the Fock matrix at each iteration leading to a third order scaling in the size of the basis set.
Linear scaling methods~\cite{Goedecker99} avoid diagonalization entirely and rely on localization properties of the system that may not generally exist in three-dimensional systems. Regardless, these methods have yet to mature to the point where they can supplant ordinary implementations.  Typical approaches to improving SCF are based on direct inversion of the iterative subspace\cite{Pulay80}, level shifting\cite{Saunders73}, quadratically convergent Newton-Raphson techniques \cite{Bacskay81}, semidefinite
programming\cite{Mazziotti14,*Mazziotti14b} 
or varying fractional occupation numbers\cite{Rabuck99} among many other approaches. The success of these approaches depends on the specific situation and the parameters chosen (e.g.~the size of the iterative subspace) and often work well in combination.  The success of these various methods has led to attempts to build black-box SCF procedures.\cite{Thogersen04,Kudin02}

Unfortunately, these methods cannot work efficiently in all cases since it was shown that Hartree-Fock is NP-complete.\cite{Schuch09,Whitfield13}  In this article, we expand upon the previous findings using translational symmetry to examine easy and hard instances when enforcing or breaking the symmetry of the underlying Hamiltonian. 
Let us note here that the 
translational invariance of the $N$-body Hamiltonian does not imply that the 
Hartree-Fock state also carries this symmetry.\cite{Overhauser60} In general, when variational 
Ans\"atze give lowest energy state without the same symmetry as the true wave function, this is called a symmetry broken solution.  The choice between the variational state with correct symmetries or the symmetry broken state is often referred to as \emph{L\"owdin's dilemma}.\cite{Lykos63} Taking this into account, we will give examples of 
1) translationally invariant HF systems that are NP-complete in the symmetry-broken space but are trivially simple when enforcing symmetry and 2) translationally invariant HF systems that are
NP-complete also when the symmetry is enforced. We will prove these NP-completeness 
results by embedding spin glass models into HF instances.

\subsection{Hartree-Fock Theory}  

A two-body Hamiltonian over $M$ sites has the general form
\begin{align}
H=&H_1+H_2\nonumber\\
=&\sum_{ij} h_{ij} a^\dag_i a^{\phantom\dagger}_j+\frac{1}{2}\sum_{ijkl} h_{ijkl} a^\dag_ia^\dag_ja^{\phantom\dagger}_ka^{\phantom\dagger}_l.\label{eq:Ham}
\end{align}
The goal of Hartree-Fock is to minimize the energy within the set $F_1$ of {single Slater} determinants 
\begin{align}
	E_{HF}&=\min_{\Psi\in F_1}\langle \Psi|H|\Psi\rangle \, .	\label{eq:HF1}
\end{align}
The single Slater determinant corresponding to the minimal value of $E_{HF}$ is called the 
Hartree-Fock state: 
\begin{equation}
\ket{\Psi_{HF}}=b_1^\dagger b_2^\dagger \cdots b_N^\dagger\ket{vac}. \label{eq:HF-state}
\end{equation} 
In this expression 
\begin{equation}
	b_i^\dag=\sum_j \ad_jC_{ji}
	\label{eq:MO}
\end{equation}
corresponds to the creation operators of the 
molecular orbitals, $\psi_i(x)=\sum_j \phi_j(x)C_{ji}$, while $\ad_j$ and $a_j$ are
creation and annihilation operators in the given atomic orbital basis, $\phi_i(x)$.
 The creation/annihilation operators must satisfy the canonical anticommutation relations 
\begin{align}
&[a_i,a_j^\dag]_+=\langle\phi_i|\phi_j\rangle\unity \, ,  \; [a_i,a_j]_+=0, \\
&[b_i,b^\dag_j ]=\delta_{i,j}\unity \, , \; \; \; \; [b_i,b_j]_+=0. \label{eq:b-anticom}
\end{align}
In other words, the \emph{orbital-rotation matrix} $C$, which connects the atomic and 
molecular orbitals, is chosen in such a way that  \eqref{eq:b-anticom} is satisfied and the
corresponding state of \eqref{eq:HF-state} is the Slater determinant that minimizes the
energy.  Let us note that if the atomic orbitals are orthogonal, $C$ will be unitary.

The \emph{charge density operator} is defined as 
\begin{equation}
	P_{pq}=\sum_i^{occ}C^{\phantom*}_{pi}C^*_{qi},
	\label{eq:P}
\end{equation}
where the summation goes only over the occupied orbitals. One can expand \eqref{eq:HF1}
using this operator as
\begin{align}
	E_{HF}&=\min_{P} \sum_{pq}P_{qp}h_{pq}+\frac12 \sum_{pqrs} P_{qp}P_{sr} (h_{prsq}-h_{prqs})\nonumber\\
	&=\min_{P}\frac12 Tr\left[P\left(F+ H_1\right)\right],
	\label{eq:Ehf}
\end{align}
where $F=F(P)=H_1+G(P)$ is the \emph{Fock operator} and $G_{pq}=G(P)_{pq}=\sum_{rs}P_{sr}(h_{prsq}-h_{prqs})$ is the \emph{mean-field potential},
which approximates the two-body interaction. Since the two-electron integrals frequently appear in
pairs, we define $A_{pqrs}=h_{pqrs}-h_{pqsr}$ as the antisymmetric integral.

\subsection{Translationally invariant fermionic systems}  
In this article, we focus our attention on translationally invariant
fermionic Hamiltonians and examine two types of Hamiltonians whose Hartree-Fock problems are both NP-complete.
Hamiltonians with translation symmetry appear in many areas of 
chemistry and physics, e.g., when modeling the electronic structure of
solids\cite{AM76}, conjugated polymers\cite{stevens1990polymer} or
fermionic atoms {in optical} lattices.\cite{bloch2008many}

The typical scenario has $L^D$ lattice sites each with 
$n_s$ orbitals per site (e.g.{,} $n_s=2$ could represent the two spin orbital associated with a spatial orbital), and then translational invariance is imposed on the sites and, finally, boundary conditions are applied in each of the $D$ dimensions. To denote translational invariance 
over multi-orbital sites requires an intra-site
label $o_i$ for  the $i$th site. In this notation, 
translational symmetry means  $h_{s_1 s_2}^{o_1 o_2}=h_{(s_1+1)( s_2+1)}^{o_1 o_2}$ 
and $h_{s_1 s_2 s_3 s_4}^{o_1 o_2 o_3 o_4} =h_{(s_1+1)( s_2+1)( s_3+1)( s_4+1)}^{o_1 o_2 o_3 o_4}$.

For simplicity, we'll consider spinless fermions ($n_s=1$) and take periodic boundary conditions, i.e., in our examples the Hamiltonians will have the symmetry property
\begin{equation*}
h_{(i+1)(j+1)}=h_{ij} \;  \;  \text{and} \; \; h_{(i+1)(j+1)(k+1)(l+1)}=h_{ijkl},
\end{equation*}
where indices are used cyclically ($k+M = k$). 
Since the system is translationally invariant, the Fourier transformed modes will often be used; these are defined over $M$ sites as 
\begin{eqnarray}
	\mathcal{F}[a]_m&=&\hat{a}_m=\frac{1}{\sqrt{M}} \sum_{x} e^{-2\pi i m x/ M }a_x \, ,\\ 
	\mathcal{F}^{-1}[\hat{a}]_x&=&a_x=\frac{1}{\sqrt{M}} \sum_{m} e^{2\pi i m x/M }\hat{a}_m \, . \label{eq:Fourier}
\end{eqnarray}

\subsection{NP-complete spin glasses}\label{sec:NPspin}
In this work, we will be {proving the NP-completeness of various Hartree-Fock problems 
by} showing that classical Ising spin glass problems can be embedded into them.\footnote{Let us note here that it is obvious that the Hartree-Fock problems are in NP, as
their solution is easy to check.}
Deciding if the ground state energy is below a certain
value for the Ising Hamiltonian,
\begin{equation}
	H_{I}= -\sum_{ij} J_{ij}S_iS_j,
	\label{eq:target}
\end{equation}
was shown to be NP-complete\cite{Barahona82} {for $J_{ij}\in\{+1,-1,0\}$} with nearest neighbor  connectivity on an $L\times L\times 2$ graph. Further investigations showed the problem for spin systems with non-planar connectivity are still NP-complete. By introducing one-body terms, even models on planar graphs were shown to be NP complete\cite{Istrail00}; and recently  various results
on three-body and higher interactions have also been published.\cite{Whitfield12,Lucas14}
 
\section{NP-complete Hartree-Fock instances with trivial translationally invariant Slater determinants}

Consider a $L\times L \times 2$ lattice, whose sites are labeled by the
integers $1, \ldots, 2L^2$ according to some ordering.
Furthermore, consider an arbitrary but fixed Ising model on this lattice 
with couplings $J_{ij} \in \{-1,0,+1\}$ when $i$ and $j$ label 
neighboring sites and  $J_{ij}=0$ otherwise. 
As discussed in Section \ref{sec:NPspin}, the ground state problem for 
this set of models is NP-{complete}.

In the following, we will define a set of translation-invariant fermionic models whose 
ground state problem can be mapped to the above Ising ground state problem, and vice versa.
{Moreover, our mapping will ensure that for each instance at least one of the ground states is a Slater determinant corresponding to the solution of the given spin glass.}
This will imply that the corresponding Hartree-Fock energy decision problem is NP-complete. 
Interestingly, it will also turn out that by restricting {the trial states} 
to {translationally} invariant 
{Slater determinants}, {the restricted HF} problem becomes trivial and thus no longer NP-complete.

We will embed an arbitrary instance of the NP-complete Ising problem on the $L\times L\times 2$ lattice to a HF problem, by constructing a fermionic embedding Hamiltonian over $M= 2 (2L^2)$ {modes} with 
the following form:
\begin{align}
H = & \sum^{M}_{i=1}\sum^{M}_{j=1}  J_{ij} 
\left(
\Ff^{\dagger}_i \Ff^{\phantom\dagger}_{M-i} +\Ff^{\dagger}_{M-i} \Ff^{\phantom\dagger}_{i}\right)
\left(  \Ff^{\dagger}_{M-j} \Ff^{\phantom\dagger}_{j}+\Ff^{\dagger}_j \Ff^{\phantom\dagger}_{M-j}  \right),
\label{eq:H}
\end{align}
where $J_{ij}$ is a fixed set of nearest neighbor couplings defining the original problem and the $\Ff_k$'s denote the Fourier transformed fermion operators, as defined in~\eqref{eq:Fourier}. Since an interaction term of form $\Ff^{\dagger}_{k_1} \Ff^{\phantom\dagger}_{k_2} \Ff^{\dagger}_{k_3} \Ff^{\phantom\dagger}_{k_4} + h.c.$ is 
translation-invariant if and only if 
\begin{equation}
k_1+k_3 - (k_2 +k_4) = 0 \mod M,
\end{equation}
the Hamiltonian is translationally invariant. 

Following the standard methods of Hartree-Fock, we 
will assume a fixed particle number. In the present case it will be 
\emph{half filling}, i.e., the number of electrons in the system
will be half the number of orbitals. However, we will work in the 
second quantized formalism (not constraining the number of
particles) and only apply the half filling condition
in the course of the proof.

\subsection{Translationally invariant Slater determinants}

Let us consider first the case when we restrict the HF problem of \eqref{eq:H} to translationally  invariant trial functions. If $|\Psi\rangle$ is a {Slater determinant}, we have by 
{Wick's theorem}~\cite{Wick50, SB2009many} 
\begin{align}
\langle \Psi | \Ff^{\dagger}_{k_1} \Ff^{\phantom\dagger}_{k_2} \Ff^{\dagger}_{k_3} \Ff^{\phantom\dagger}_{k_4}  |\Psi\rangle &= 
\langle \Psi | \Ff^{\dagger}_{k_1} \Ff^{\phantom\dagger}_{k_2} |\Psi\rangle 
\langle \Psi | \Ff^{\dagger}_{k_3} \Ff^{\phantom\dagger}_{k_4}  |\Psi\rangle \nonumber \\
&- \langle \Psi | \Ff^{\dagger}_{k_1} \Ff^{\phantom\dagger}_{k_4} |\Psi\rangle 
\langle \Psi | \Ff^{\dagger}_{k_3} \Ff^{\phantom\dagger}_{k_2}  |\Psi\rangle.
\end{align} 
Furthermore, if $|\Psi\rangle$ is a translationally {invariant},
then $\langle \Psi | \Ff^{\dagger}_{p} \Ff^{\phantom\dagger}_{q} |\Psi\rangle =0$
if $p \ne q$. These two statements imply that for all translationally invariant
{Slater determinants}, with or without half filling, $\langle \Psi | H  |\Psi\rangle=0$. Hence, this restricted HF ground state problem is trivial.

\subsection{Broken symmetry Slater determinants}

Next we will show that, contrary to the previous case, the \emph{non-restricted} 
Hartree-Fock problem is NP-complete. 
For this, let us observe that an alternative characterization of \eqref{eq:H} is given by
\begin{align}
H=&\sum_{i,j} J_{ij}  \left[(n^{(+)}_i-n^{(-)}_i) (n^{(+)}_j -  n^{(-)}_j) \right] , \label{eq:H2_alt}
\end{align}
where $n^{(\pm)}_k={c^{(\pm)}_k}^\dag c^{(\pm)}_k$ and $c^{(\pm)}_i=(\Ff_i \pm \Ff_{-i})/\sqrt{2}$. Observe that the gerade $(+)$ and ungerade $(-)$ orbitals are orthogonal, so it
follows directly from the fermionic algebra that $[c_m^{(+)},(c_n^{(-)})^\dag]_+=0$. 

Let us  note here that the terms $(n^{(+)}_i-n^{(-)}_i)$ in \eqref{eq:H2_alt} are similar to those found in the orbital pair pseudo-spin mapping originally used to prove that Hartree-Fock is NP-complete.\cite{Schuch09} However, the present mapping differs in an important way: unlike the previous construction, we do not need an additional penalty term in the Hamiltonian to enforce the orbital pairing 
condition, as will be clear 
from the discussion below.

Since the $(+)$ and $(-)$ orbitals are orthogonal, we can immediately write all $2^{2M}$ eigenstates of \eqref{eq:H2_alt} (and hence of \eqref{eq:H}), as
\begin{equation}
\ket{s}=\prod_{i=1}^M  ({c^{(+)}_i}^\dag)^{\sigma_i}  ({c^{(-)}_i}^\dag)^{\tau_i} |vac\rangle, \label{eq:eigenvec}
\end{equation}
with energies,
\begin{equation}
\bra{s}H\ket{s}=\sum_{i,j}J_{ij}(\sigma_i\sigma_j+\tau_i\tau_j-\sigma_i\tau_j-\sigma_i\tau_j).
\label{eq:energy}
\end{equation}
Since the modes $c^{(\pm)}_i$ do not carry the underlying translational symmetry of the Hamiltonian, neither will the eigenstates $|s\rangle$ inherit this symmetry 
(except for some degenerate cases as the vacuum and the completely filled state).

The ground state energy of \eqref{eq:H} is thus the minimum value attained by \eqref{eq:energy}.  To demonstrate that this minimum and the ground state energy of the Ising spin glass of \eqref{eq:target} are in correspondence, we will first give a lower bound on \eqref{eq:energy}, which relies on bipartite graph properties, and then we show that the \emph{orbital paired state} corresponding to the solution of the Ising spin glass has exactly an energy saturating this lower bound. 

As the $L \times L \times 2$ rectangular lattice is a bipartite lattice, one can label alternating spins even and odd such that no two spins with the same label share interactions.  The labeling then corresponds to a bipartition of the lattice.  The relevant property of bipartite lattices is that the energy spectrum of an Ising spin glass is symmetric.  To prove that the largest energy, $E_{max}$ is equal to $-E_{min}$, we will investigate the lowest energy configuration of $-H_I$ instead of the highest energy of $H_I$. If $s=[s_1,...,s_M]$ is a minimal configuration for Hamiltonian $H_I$, then given a bipartition, setting $s_i'= -s_i$ for sites with the even label and $s_i'=s_i$ for sites $i$ with odd label, we obtain a ground state $s'$ for $-H_I$ with energy $E_{min}$. This implies that $E_{max}=-E_{min}$ as we intended. 

In \eqref{eq:energy}, there are four terms. The first two have minimal energy $E_{min}$ and the second two have energy at least $-E_{max}$. It follows, using $E_{max}=-E_{min}$, that the minimal energy configuration of \eqref{eq:energy} is at least $4E_{min}$, i.e.
\begin{equation}
\bra{s}H\ket{s} \le 4 E_{min}.
\end{equation}

Now, let again $s=[s_1,...,s_M]$ denote the minimal configuration for the Ising Hamiltonian $H_I$, and 
consider an eigenstate of \eqref{eq:H} using the orbital pair scheme as
\begin{equation}
\prod_{i=1}^M  ({c^{(+)}_i}^\dag)^{\sigma_i}  ({c^{(-)}_i}^\dag)^{1-\sigma_i} |vac\rangle, 
\end{equation}
with $\sigma_i=(s_i-1)/2$. This state is at half-filling and, due to the bipartite
nature of the underlying lattice, it attains the energy $4E_{min}$, thus this must be a ground state.  This means that the Hartree-Fock energy problem for \eqref{eq:H} is equivalent to finding the lowest energy for an Ising spin glass, and hence it is NP-complete.

\section{Mapping NP-complete spin systems to translationally invariant Hartree-Fock instances} 

\begin{figure}
	\begin{center}
		\includegraphics[scale=0.57]{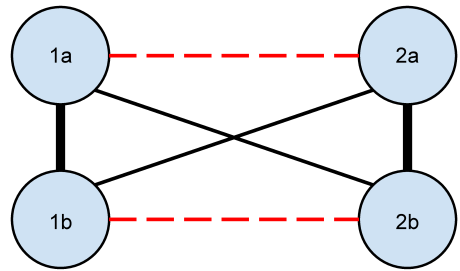}
	\end{center}
	\caption{(Color online) An Ising gadget used to convert a given Ising problem into a new Ising problem with twice as many spins whose solution all have zero magnetization.  The solutions of the problems in the standard basis are in {one-to-one} correspondence. This is achieved by appropriately coupling pairs of spins where each pair represents a spin from the original problem.  Depicted is the appropriate couplings in the new graph if spins $i$ and $j$ were connected with coupling strength $J$. The pairing coupling is given by $C=4d_{max}J_{max}$ with $d_{max}$ the maximum degree and $J_{max}$ the largest coupling in the given graph.}
	\label{fig:IsingGadget}
\end{figure}

{We will now turn to a set of Hamiltonians where the Hartree-Fock problem is NP-complete even when we restrict ourselves to translationally invariant Slater determinants.} 

\subsection{Gadgets enforcing half filling}

We will again like to assume a fixed particle number, more concretely half filling. 
As the usual embedding of spins into fermions maps the number of electrons to number of 
`up' spins, we  have to build up Ising models where the ground state has 
zero magnetization. In the present example this will be done by introducing an Ising gadget
similar to the pair-orbital construction used to map spin systems to fermionic 
systems.\cite{Schuch09,Whitfield13}  The idea is to fix the zero magnetization  by introducing pairs of spins $j\alpha$ and $j\beta$ to represent single spin $j$ in 
the original Ising spin glass system.  Strong anti-ferromagnetic coupling between $j\alpha$ and $j\beta$ 
enforce anti-parallel alignment. To ensure that the mapped
spins are correctly paired, selecting the pair coupling strength, $\mathcal{C}$, to be
 four times the maximum degree, $d_{max}$, of a spin is sufficient.
If $j=+1$, then we should have $(j\alpha,j\beta)=(+1,-1)$ and, if $j=-1$, then $(j\alpha,j\beta)=(-1,+1)$.
Coupling, $J_{ij}$, between two spins $i$ and $j$ is effected by letting
{\[
J_{i\beta,j\beta}=J_{i\alpha,j\alpha}=-J_{i\alpha,j\beta}=-J_{i\beta,j\alpha}=J_{ij} \, .
\]}
This gadget is depicted in Fig.~\ref{fig:IsingGadget}. The energy of the new Hamiltonian, $E'$, is
{\[
E'=4E-N\mathcal{C}.
\]}

\subsection{NP-completeness}

Now we show that there is an 
embedding of arbitrary NP-complete Ising problems 
into instances of the Hartree-Fock method applied to translationally invariant system.  In other words, 
we wish to design a mapping from a given set of couplings 
$\left\{J_{ij}\in\mathcal{Z}\right\}$ to the sets 
$\{h_{ij}: h_{ij}=h_{i+1,j+1}\}$ and $\{h_{ijkl}: h_{ijkl}=h_{i+1,j+1,k+1,l+1}\}$ 
defining a one- and two-body fermionic Hamiltonian with mean-field energy equal to the
energy of Ising system $H(\{J_{ij}\})$.

To begin, assume that $\{J_{ij}\}$ has been fixed and the total magnetization is zero.  We now
consider the case where there are $M$ orbitals and $N=M/2$ fermions in the system.  Returning 
to \eqref{eq:P}, we expand the summation over all basis functions and use an indicator vector
$R$ where $R_i=1$ if orbital $i$ is occupied and zero otherwise. Since the Ising spins take
values $\pm1$, we convert from $R$ to $S$ using $R=(S+1)/2$. Putting this together,
{\begin{equation}
	P_{pq}=\sum_{i=1}^M C_{pi}R_iC_{qi}^*=\frac12 \sum_{i=1}^M S_iC_{pi}C_{qi}^*+
\frac12\delta_{pq} .
\end{equation}}
We have assumed that the atomic orbital basis is orthogonal and used the fact that $CC^\dagger=\mathbf{1}$. Next, we express the Hartree-Fock energy in terms
of $S_i=\pm1$ by substituting into \eqref{eq:Ehf}{:}
{\begin{eqnarray}
E(P)  	&=&-\sum_{ij} \left[\frac{1}8\sum_{pqrs}C_{pi}^*C_{qi}C_{rj}^*C_{sj}A_{prqs}\right]S_iS_j\nonumber\\
	&& +\sum_j \left[\frac14\sum_{pq} C_{qj}C_{pj}^*\left(2h_{pq}+\sum_{r}A_{prrq}\right)\right]S_j\nonumber\\
	&& +\frac12 \sum_{p}h_{pp}+\frac18\sum_{pr}A_{prrp} \, .
\end{eqnarray}}
To obtain the energy function of $H(\{J_{ij}\})=-\sum_{ij} J_{ij}S_iS_j$, we would like the single spin 
Hamiltonian to be zero so we define $h_{pq}:=-\sum_{r}A_{prrq}/2$.

Before enforcing equality for the two-spin interactions, we make use of the translational invariance 
of the Hamiltonian. 
{If we restrict the trial states to translationally invariant Slater determinants,
the fact that $H_1$ possesses translation symmetry implies that the mean-field potential 
is also translation-invariant:
$G_{pq}= \sum_{rs}P_{sr}A_{prsq}=\sum_{rs}P_{s+1, r+1}A_{p+1,r+1,s+1,q+1} = G_{p+1, q+1}$.
}
As a result, the 
eigenvectors of the Fock matrix are given by
\begin{equation}
\left|\hat{k}\right\rangle=\frac{1}{\sqrt{M}}\sum_{x={1}}^M e^{- 2\pi ik x/M}\ket{x} \, {.} 
\label{eq:eigen_fourier}
\end{equation}
Thus,  $F$ is brought to diagonal form by $C_{kn}=\exp(-i2\pi kn/M)/\sqrt M$ as
$CFC^\dag=\Lambda$.

Despite knowing the eigendecomposition, selecting the correct orbitals is still NP-hard. 
We prove this statement by equating $J_{ij}$ from the Ising problem to the two-body
interaction of the fermionic system $J_{ij}=\sum_{pqrs}C_{pi}^*C_{qi}C_{rj}^*C_{sj}A_{prsq}/8$.
We utilize the Fourier transform to get the appropriate form and account for the 
anti-symmetry of the $A$ explicitly,
{\begin{eqnarray}
	A_{pqrs}&=& -\frac{8}{M}\sum_{tu}C_{pt}C_{qu} (C_{ru}C_{st}-C_{su}C_{rt})^*J_{tu}.
		\label{eq:ftV}
\end{eqnarray}}
A simple calculation verifies that the inverse Fourier transform of $A_{pqrs}$ with pairs $(m,p-s)$ 
and $(n,q-r)$ leads to $J_{mn}$ as desired.

After setting $h_{pq}=-\sum_{r}A_{prrq}/2$, we have a fermionic Hamiltonian with Hartree-Fock energy 
\begin{equation}
	E_{HF}(S)=H_{I}(S)+\sum_{mn}J_{mn}.
\end{equation}
This mapping between the two problems requires polynomial time overhead implying that the Hartree-Fock for 
translationally invariant systems is NP-complete.

\subsection{Numerical considerations}

\begin{figure}[tb]
        \begin{center}
		\includegraphics[scale=0.37]{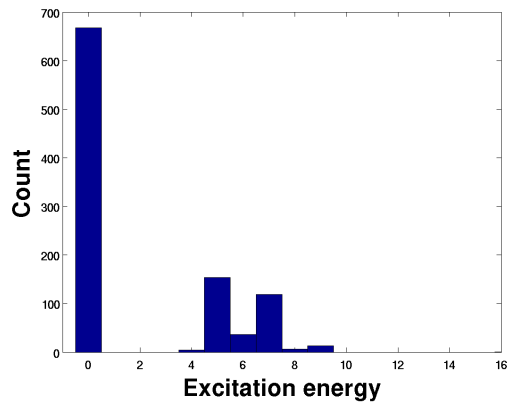}
        \end{center}
	\caption{Histogram of {excitation} energies found using the Hartree-Fock algorithm on 1000 instances derived from the spin glass corresponding to the coupling matrix of \eqref{eq:exJ}.}
	\label{fig:hist}
\end{figure}

It is known that the commutator of the charge density and  Fock operators is zero, $[P,F]=0$, at local minima including the true Hartree-Fock global minimum.  For this reason, the 
direct inversion iterative subspace method~\cite{Pulay80} utilizes the norm of the $[P,F]$ commutator to accelerate convergence.  However, in our model, all the translationally invariant states are local minima and thus the commutator is always zero.  While this is no longer a useful error measure, other measures such as the energy based direct inversion iterative scheme~\cite{Kudin02} can still be useful.

\begin{figure*}[t]
	\begin{center}
                \includegraphics[scale=0.23]{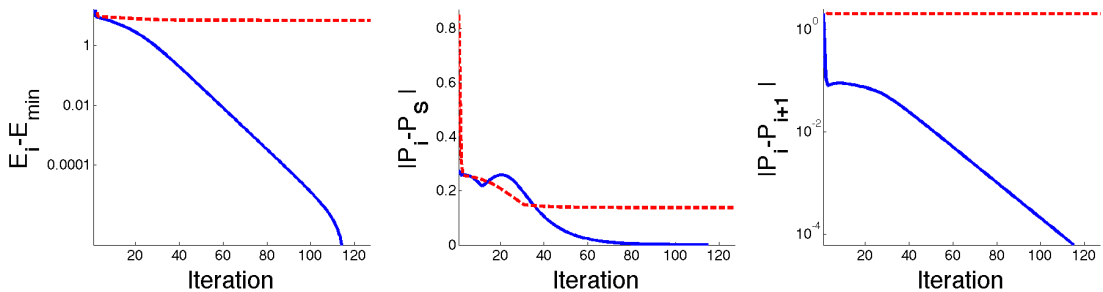}
	\end{center}
	\caption{Comparison of successful (solid blue) and unsuccessful (dashed red) self-consistent field method runs.  The unsuccessful run converged to $E=-5$ instead of $E=-10$ in the successful run.  In the leftmost plot, the convergence to the final energy is shown to be exponential in both cases.  The middle plot is the norm of $P-SPS^\dag$ where $S$ is the shift operator.  It shows that as the algorithm progresses, the SCF method finds the translationally invariant subspace. Finally, in the last plot, the change in norm between subsequent iterations is plotted.  In the non-convergent case, the distance between steps stays approximately two but decays slightly after each iteration. }
	\label{fig:3}
\end{figure*}

To illustrate the mapping described, we explore the utility of the Hartree-Fock self-consistent formulation on a test spin glass with coupling matrix 
\begin{equation}
	J=\left[
		\begin{array}{rrrrrr}
   0 &  -1 & -1  & -1  &  0&   -1\\
      &   0  &  1  &  0  &  1&    1\\
      &       &  0  &  1  &  1&    0\\
      &       &      &  0  & \phantom{-}1&    1\\
      &       &     &       &  0&   -1\\
      &       &     &       &   &    0
	\end{array}
		\right].
		\label{eq:exJ}
\end{equation}
The spectrum of the spin Hamiltonian is $\{-10,-6,-2,2,6\}$ and the system has 
ground states ${s}=[-1,1,1,1,1,1]$ and $-{s}$. After converting this to an instance of Hartree-Fock, we examined 1000 runs of the basic self-consistent algorithm limited to 128 cycles. Beginning from Haar {distributed} charge density {matrices}, we found that the algorithm converged on 570 of 1000 instances and an additional 98 instances found the ground state energy despite failing to converge within 128 cycles. A histogram of the results is shown in figure~\ref{fig:hist}.
We provide also an example of converged and unconverged instances in figure \ref{fig:3}.

\section{Conclusion}

We continued the line of inquiry of Hamiltonian complexity~\cite{Osborne12} in 
the context of chemistry\cite{Whitfield13} with a study of the complexity of 
the translationally invariant Hartree-Fock problem, proving that it is NP-complete. 
It is worth pointing out that our results utilize highly spatial non-local Hamiltonians since
the models are local in Fourier space. The non-locality is likely to be required 
in order to allow enough parameters for the models to be both NP-hard and translationally 
invariant. However, there are some hardness results in this direction for local and translationally 
invariant systems.\cite{Gottesman13}

This work is the first in a series of inquiries aiming to understand the appearance of 
difficult instances of the Hartree-Fock and post-Hartree-Fock 
algorithms.  With the mappings provided here, we also open the door to using the 
self-consistent method in the study of spin glasses. The implications of computer 
science in chemistry has yet to be fully explored; these studies, together with other 
recent work, are the first steps in this direction.

\section*{Acknowledgements}
We would like to thank A. Pagani and A. Ramezanpour for inspiring this work as well as ISI where 
parts of this work were completed. We also acknowledge helpful discussions with F. Verstraete, Z. Pusk\'as and J. D. Biamonte.  JDW thanks the VCQ and Ford fellowships for support, and ZZ acknowledges 
funding by the British Engineering and Physical Sciences Research Council (EPSRC), the 
Basque Government (Project No. IT4720-10) and by the European Union through the ERC Starting 
Grant GEDENTQOPT and the CHIST-ERA QUASAR project.\\


\begin{thebibliography}{27}%
\makeatletter
\providecommand \@ifxundefined [1]{%
 \@ifx{#1\undefined}
}%
\providecommand \@ifnum [1]{%
 \ifnum #1\expandafter \@firstoftwo
 \else \expandafter \@secondoftwo
 \fi
}%
\providecommand \@ifx [1]{%
 \ifx #1\expandafter \@firstoftwo
 \else \expandafter \@secondoftwo
 \fi
}%
\providecommand \natexlab [1]{#1}%
\providecommand \enquote  [1]{``#1''}%
\providecommand \bibnamefont  [1]{#1}%
\providecommand \bibfnamefont [1]{#1}%
\providecommand \citenamefont [1]{#1}%
\providecommand \href@noop [0]{\@secondoftwo}%
\providecommand \href [0]{\begingroup \@sanitize@url \@href}%
\providecommand \@href[1]{\@@startlink{#1}\@@href}%
\providecommand \@@href[1]{\endgroup#1\@@endlink}%
\providecommand \@sanitize@url [0]{\catcode `\\12\catcode `\$12\catcode
  `\&12\catcode `\#12\catcode `\^12\catcode `\_12\catcode `\%12\relax}%
\providecommand \@@startlink[1]{}%
\providecommand \@@endlink[0]{}%
\providecommand \url  [0]{\begingroup\@sanitize@url \@url }%
\providecommand \@url [1]{\endgroup\@href {#1}{\urlprefix }}%
\providecommand \urlprefix  [0]{URL }%
\providecommand \Eprint [0]{\href }%
\providecommand \doibase [0]{http://dx.doi.org/}%
\providecommand \selectlanguage [0]{\@gobble}%
\providecommand \bibinfo  [0]{\@secondoftwo}%
\providecommand \bibfield  [0]{\@secondoftwo}%
\providecommand \translation [1]{[#1]}%
\providecommand \BibitemOpen [0]{}%
\providecommand \bibitemStop [0]{}%
\providecommand \bibitemNoStop [0]{.\EOS\space}%
\providecommand \EOS [0]{\spacefactor3000\relax}%
\providecommand \BibitemShut  [1]{\csname bibitem#1\endcsname}%
\let\auto@bib@innerbib\@empty
\bibitem [{\citenamefont {Hartree}(1928)}]{Hartree28}%
  \BibitemOpen
  \bibfield  {author} {\bibinfo {author} {\bibfnamefont {D.}~\bibnamefont
  {Hartree}},\ }\href@noop {} {\bibfield  {journal} {\bibinfo  {journal} {Proc.
  Camb. Phil. Soc.}\ }\textbf {\bibinfo {volume} {24}},\ \bibinfo {pages} {89}
  (\bibinfo {year} {1928})}\BibitemShut {NoStop}%
\bibitem [{\citenamefont {Fock}(1930)}]{Fock30}%
  \BibitemOpen
  \bibfield  {author} {\bibinfo {author} {\bibfnamefont {V.}~\bibnamefont
  {Fock}},\ }\href@noop {} {\bibfield  {journal} {\bibinfo  {journal} {Z.
  Phys.}\ }\textbf {\bibinfo {volume} {61}},\ \bibinfo {pages} {723} (\bibinfo
  {year} {1930})}\BibitemShut {NoStop}%
\bibitem [{\citenamefont {Goedecker}(1999)}]{Goedecker99}%
  \BibitemOpen
  \bibfield  {author} {\bibinfo {author} {\bibfnamefont {S.}~\bibnamefont
  {Goedecker}},\ }\href@noop {} {\bibfield  {journal} {\bibinfo  {journal}
  {Rev. Mod. Phys.}\ }\textbf {\bibinfo {volume} {71}},\ \bibinfo {pages}
  {1085} (\bibinfo {year} {1999})}\BibitemShut {NoStop}%
\bibitem [{\citenamefont {Pulay}(1980)}]{Pulay80}%
  \BibitemOpen
  \bibfield  {author} {\bibinfo {author} {\bibfnamefont {P.}~\bibnamefont
  {Pulay}},\ }\href@noop {} {\bibfield  {journal} {\bibinfo  {journal} {Chem.
  Phys. Lett.}\ }\textbf {\bibinfo {volume} {73}},\ \bibinfo {pages} {393}
  (\bibinfo {year} {1980})}\BibitemShut {NoStop}%
\bibitem [{\citenamefont {Saunders}\ and\ \citenamefont
  {Hillier}(1973)}]{Saunders73}%
  \BibitemOpen
  \bibfield  {author} {\bibinfo {author} {\bibfnamefont {V.~R.}\ \bibnamefont
  {Saunders}}\ and\ \bibinfo {author} {\bibfnamefont {I.~H.}\ \bibnamefont
  {Hillier}},\ }\href@noop {} {\bibfield  {journal} {\bibinfo  {journal} {Intl.
  J. Quant. Chem.}\ }\textbf {\bibinfo {volume} {7}},\ \bibinfo {pages} {699}
  (\bibinfo {year} {1973})}\BibitemShut {NoStop}%
\bibitem [{\citenamefont {Bacskay}(1981)}]{Bacskay81}%
  \BibitemOpen
  \bibfield  {author} {\bibinfo {author} {\bibfnamefont {G.~B.}\ \bibnamefont
  {Bacskay}},\ }\href@noop {} {\bibfield  {journal} {\bibinfo  {journal} {Chem.
  Phys.}\ }\textbf {\bibinfo {volume} {61}},\ \bibinfo {pages} {385} (\bibinfo
  {year} {1981})}\BibitemShut {NoStop}%
\bibitem [{\citenamefont {Veeraraghavan}\ and\ \citenamefont
  {Mazziotti}(2014{\natexlab{a}})}]{Mazziotti14}%
  \BibitemOpen
  \bibfield  {author} {\bibinfo {author} {\bibfnamefont {S.}~\bibnamefont
  {Veeraraghavan}}\ and\ \bibinfo {author} {\bibfnamefont {D.~A.}\ \bibnamefont
  {Mazziotti}},\ }\href@noop {} {\bibfield  {journal} {\bibinfo  {journal} {J.
  Chem. Phys.}\ }\textbf {\bibinfo {volume} {140}},\ \bibinfo {pages} {124106}
  (\bibinfo {year} {2014}{\natexlab{a}})}\BibitemShut {NoStop}%
\bibitem [{\citenamefont {Veeraraghavan}\ and\ \citenamefont
  {Mazziotti}(2014{\natexlab{b}})}]{Mazziotti14b}%
  \BibitemOpen
  \bibfield  {author} {\bibinfo {author} {\bibfnamefont {S.}~\bibnamefont
  {Veeraraghavan}}\ and\ \bibinfo {author} {\bibfnamefont {D.~A.}\ \bibnamefont
  {Mazziotti}},\ }\href@noop {} {\bibfield  {journal} {\bibinfo  {journal}
  {Phys. Rev. A}\ }\textbf {\bibinfo {volume} {89}},\ \bibinfo {pages}
  {010502(R)} (\bibinfo {year} {2014}{\natexlab{b}})}\BibitemShut {NoStop}%
\bibitem [{\citenamefont {Rabuck}\ and\ \citenamefont
  {Scuseria}(1999)}]{Rabuck99}%
  \BibitemOpen
  \bibfield  {author} {\bibinfo {author} {\bibfnamefont {A.~D.}\ \bibnamefont
  {Rabuck}}\ and\ \bibinfo {author} {\bibfnamefont {G.~E.}\ \bibnamefont
  {Scuseria}},\ }\href@noop {} {\bibfield  {journal} {\bibinfo  {journal} {J.
  Chem. Phys.}\ }\textbf {\bibinfo {volume} {110}},\ \bibinfo {pages} {695}
  (\bibinfo {year} {1999})}\BibitemShut {NoStop}%
\bibitem [{\citenamefont {Th{\o}gersen}\ \emph {et~al.}(2004)\citenamefont
  {Th{\o}gersen}, \citenamefont {Olsen}, \citenamefont {Yeager}, \citenamefont
  {J{\o}rgensen}, \citenamefont {Salek},\ and\ \citenamefont
  {Helgaker}}]{Thogersen04}%
  \BibitemOpen
  \bibfield  {author} {\bibinfo {author} {\bibfnamefont {L.}~\bibnamefont
  {Th{\o}gersen}}, \bibinfo {author} {\bibfnamefont {J.}~\bibnamefont {Olsen}},
  \bibinfo {author} {\bibfnamefont {D.}~\bibnamefont {Yeager}}, \bibinfo
  {author} {\bibfnamefont {P.}~\bibnamefont {J{\o}rgensen}}, \bibinfo {author}
  {\bibfnamefont {P.}~\bibnamefont {Salek}}, \ and\ \bibinfo {author}
  {\bibfnamefont {T.}~\bibnamefont {Helgaker}},\ }\href@noop {} {\bibfield
  {journal} {\bibinfo  {journal} {J. Chem. Phys.}\ }\textbf {\bibinfo {volume}
  {121}},\ \bibinfo {pages} {16} (\bibinfo {year} {2004})}\BibitemShut
  {NoStop}%
\bibitem [{\citenamefont {Kudin}, \citenamefont {Scuseria},\ and\ \citenamefont
  {Canc\`es}(2002)}]{Kudin02}%
  \BibitemOpen
  \bibfield  {author} {\bibinfo {author} {\bibfnamefont {K.~N.}\ \bibnamefont
  {Kudin}}, \bibinfo {author} {\bibfnamefont {G.~E.}\ \bibnamefont {Scuseria}},
  \ and\ \bibinfo {author} {\bibfnamefont {E.}~\bibnamefont {Canc\`es}},\
  }\href@noop {} {\bibfield  {journal} {\bibinfo  {journal} {J. Chem. Phys.}\
  }\textbf {\bibinfo {volume} {116}},\ \bibinfo {pages} {8255} (\bibinfo {year}
  {2002})}\BibitemShut {NoStop}%
\bibitem [{\citenamefont {Schuch}\ and\ \citenamefont
  {Verstraete}(2009)}]{Schuch09}%
  \BibitemOpen
  \bibfield  {author} {\bibinfo {author} {\bibfnamefont {N.}~\bibnamefont
  {Schuch}}\ and\ \bibinfo {author} {\bibfnamefont {F.}~\bibnamefont
  {Verstraete}},\ }\href@noop {} {\bibfield  {journal} {\bibinfo  {journal}
  {Nature Phys.}\ }\textbf {\bibinfo {volume} {5}},\ \bibinfo {pages} {732}
  (\bibinfo {year} {2009})},\ \bibinfo {note} {also see appendix of
  arxiv:0712.0483}\BibitemShut {NoStop}%
\bibitem [{\citenamefont {Whitfield}, \citenamefont {Love},\ and\ \citenamefont
  {Aspuru-Guzik}(2013)}]{Whitfield13}%
  \BibitemOpen
  \bibfield  {author} {\bibinfo {author} {\bibfnamefont {J.~D.}\ \bibnamefont
  {Whitfield}}, \bibinfo {author} {\bibfnamefont {P.~J.}\ \bibnamefont {Love}},
  \ and\ \bibinfo {author} {\bibfnamefont {A.}~\bibnamefont {Aspuru-Guzik}},\
  }\href@noop {} {\bibfield  {journal} {\bibinfo  {journal} {Phys. Chem. Chem.
  Phys.}\ }\textbf {\bibinfo {volume} {15}},\ \bibinfo {pages} {397} (\bibinfo
  {year} {2013})}\BibitemShut {NoStop}%
\bibitem [{\citenamefont {Overhauser}(1960)}]{Overhauser60}%
  \BibitemOpen
  \bibfield  {author} {\bibinfo {author} {\bibfnamefont {A.~W.}\ \bibnamefont
  {Overhauser}},\ }\href@noop {} {\bibfield  {journal} {\bibinfo  {journal}
  {Phys. Rev. Lett.}\ }\textbf {\bibinfo {volume} {4}},\ \bibinfo {pages} {415}
  (\bibinfo {year} {1960})}\BibitemShut {NoStop}%
\bibitem [{\citenamefont {Lykos}\ and\ \citenamefont {Pratt}(1963)}]{Lykos63}%
  \BibitemOpen
  \bibfield  {author} {\bibinfo {author} {\bibfnamefont {P.}~\bibnamefont
  {Lykos}}\ and\ \bibinfo {author} {\bibfnamefont {G.~W.}\ \bibnamefont
  {Pratt}},\ }\href@noop {} {\bibfield  {journal} {\bibinfo  {journal} {Rev.
  Mod. Phys.}\ }\textbf {\bibinfo {volume} {35}},\ \bibinfo {pages} {496}
  (\bibinfo {year} {1963})}\BibitemShut {NoStop}%
\bibitem [{\citenamefont {Ashcroft}\ and\ \citenamefont {Mermin}(1976)}]{AM76}%
  \BibitemOpen
  \bibfield  {author} {\bibinfo {author} {\bibfnamefont {N.~W.}\ \bibnamefont
  {Ashcroft}}\ and\ \bibinfo {author} {\bibfnamefont {N.~D.}\ \bibnamefont
  {Mermin}},\ }\href@noop {} {\emph {\bibinfo {title} {Solid State Physics}}}\
  (\bibinfo  {publisher} {Thomson learning},\ \bibinfo {year}
  {1976})\BibitemShut {NoStop}%
\bibitem [{\citenamefont {Stevens}(1990)}]{stevens1990polymer}%
  \BibitemOpen
  \bibfield  {author} {\bibinfo {author} {\bibfnamefont {M.~P.}\ \bibnamefont
  {Stevens}},\ }\href@noop {} {\emph {\bibinfo {title} {Polymer chemistry}}}\
  (\bibinfo  {publisher} {Oxford University Press, New York},\ \bibinfo {year}
  {1990})\BibitemShut {NoStop}%
\bibitem [{\citenamefont {Bloch}, \citenamefont {Dalibard},\ and\ \citenamefont
  {Zwerger}(2008)}]{bloch2008many}%
  \BibitemOpen
  \bibfield  {author} {\bibinfo {author} {\bibfnamefont {I.}~\bibnamefont
  {Bloch}}, \bibinfo {author} {\bibfnamefont {J.}~\bibnamefont {Dalibard}}, \
  and\ \bibinfo {author} {\bibfnamefont {W.}~\bibnamefont {Zwerger}},\
  }\href@noop {} {\bibfield  {journal} {\bibinfo  {journal} {Rev. Mod. Phys.}\
  }\textbf {\bibinfo {volume} {80}},\ \bibinfo {pages} {885} (\bibinfo {year}
  {2008})}\BibitemShut {NoStop}%
\bibitem [{Note1()}]{Note1}%
  \BibitemOpen
  \bibinfo {note} {Let us note here that it is obvious that the Hartree-Fock
  problems are in NP, as their solution is easy to check.}\BibitemShut {Stop}%
\bibitem [{\citenamefont {Barahona}(1982)}]{Barahona82}%
  \BibitemOpen
  \bibfield  {author} {\bibinfo {author} {\bibfnamefont {F.}~\bibnamefont
  {Barahona}},\ }\href@noop {} {\bibfield  {journal} {\bibinfo  {journal} {J.
  Phys. A: Math. Gen.}\ }\textbf {\bibinfo {volume} {15}},\ \bibinfo {pages}
  {3241} (\bibinfo {year} {1982})}\BibitemShut {NoStop}%
\bibitem [{\citenamefont {Istrail}(2000)}]{Istrail00}%
  \BibitemOpen
  \bibfield  {author} {\bibinfo {author} {\bibfnamefont {S.}~\bibnamefont
  {Istrail}},\ }\href@noop {} {\bibfield  {journal} {\bibinfo  {journal} {Proc.
  32nd ACM Symp. on Theory of Comp. (STOC '00)}\ ,\ \bibinfo {pages} {87}}
  (\bibinfo {year} {2000})}\BibitemShut {NoStop}%
\bibitem [{\citenamefont {Whitfield}, \citenamefont {Faccin},\ and\
  \citenamefont {Biamonte}()}]{Whitfield12}%
  \BibitemOpen
  \bibfield  {author} {\bibinfo {author} {\bibfnamefont {J.~D.}\ \bibnamefont
  {Whitfield}}, \bibinfo {author} {\bibfnamefont {M.}~\bibnamefont {Faccin}}, \
  and\ \bibinfo {author} {\bibfnamefont {J.~D.}\ \bibnamefont {Biamonte}},\
  }\href@noop {} {\bibfield  {journal} {\bibinfo  {journal} {EPL}\ }\textbf
  {\bibinfo {volume} {99}},\ \bibinfo {pages} {57004}}\BibitemShut {NoStop}%
\bibitem [{\citenamefont {Lucas}(2014)}]{Lucas14}%
  \BibitemOpen
  \bibfield  {author} {\bibinfo {author} {\bibfnamefont {A.}~\bibnamefont
  {Lucas}},\ }\href@noop {} {\bibfield  {journal} {\bibinfo  {journal} {Front.
  Physics}\ }\textbf {\bibinfo {volume} {2}},\ \bibinfo {pages} {5} (\bibinfo
  {year} {2014})},\ \bibinfo {note} {10.3389/fphy.2014.00005}\BibitemShut
  {NoStop}%
\bibitem [{\citenamefont {Wick}(1950)}]{Wick50}%
  \BibitemOpen
  \bibfield  {author} {\bibinfo {author} {\bibfnamefont {G.~C.}\ \bibnamefont
  {Wick}},\ }\href@noop {} {\bibfield  {journal} {\bibinfo  {journal} {Phys.
  Rev.}\ }\textbf {\bibinfo {volume} {80}},\ \bibinfo {pages} {268} (\bibinfo
  {year} {1950})}\BibitemShut {NoStop}%
\bibitem [{\citenamefont {Shavitt}\ and\ \citenamefont
  {Bartlett}(2009)}]{SB2009many}%
  \BibitemOpen
  \bibfield  {author} {\bibinfo {author} {\bibfnamefont {I.}~\bibnamefont
  {Shavitt}}\ and\ \bibinfo {author} {\bibfnamefont {R.~J.}\ \bibnamefont
  {Bartlett}},\ }\href@noop {} {\emph {\bibinfo {title} {Many-body methods in
  chemistry and physics: MBPT and coupled-cluster theory}}}\ (\bibinfo
  {publisher} {Cambridge University Press},\ \bibinfo {year}
  {2009})\BibitemShut {NoStop}%
\bibitem [{\citenamefont {Osborne}(2012)}]{Osborne12}%
  \BibitemOpen
  \bibfield  {author} {\bibinfo {author} {\bibfnamefont {T.~J.}\ \bibnamefont
  {Osborne}},\ }\href@noop {} {\bibfield  {journal} {\bibinfo  {journal} {Rep.
  Prog. Phys.}\ }\textbf {\bibinfo {volume} {75}},\ \bibinfo {pages} {022001}
  (\bibinfo {year} {2012})}\BibitemShut {NoStop}%
\bibitem [{\citenamefont {Gottesman}\ and\ \citenamefont
  {Irani}(2013)}]{Gottesman13}%
  \BibitemOpen
  \bibfield  {author} {\bibinfo {author} {\bibfnamefont {D.}~\bibnamefont
  {Gottesman}}\ and\ \bibinfo {author} {\bibfnamefont {S.}~\bibnamefont
  {Irani}},\ }\href@noop {} {\bibfield  {journal} {\bibinfo  {journal} {Theory
  OF Computing}\ }\textbf {\bibinfo {volume} {9}},\ \bibinfo {pages} {31}
  (\bibinfo {year} {2013})},\ \bibinfo {note} {{arXiv}:0905.2419}\BibitemShut
  {NoStop}%
\end{thebibliography}
\end{document}